\documentclass[%
preprint,
 amsmath,amssymb,
 aps,
]{revtex4-1}

\usepackage{color}
\usepackage{graphicx}
\usepackage{dcolumn}
\usepackage{bm}

\def\sub#1{_{\mathrm{#1}}}
\def\up#1{^{\mathrm{#1}}}
\def\Vec#1{\boldsymbol #1}

\newcommand {\beq}{\begin{eqnarray}}
\newcommand {\eeq}{\end{eqnarray}}

\begin{document}


\title{
Non-relativistic Nambu-Goldstone modes \\
propagating along a skyrmion line 
}

\author{Michikazu Kobayashi$^1$, Muneto Nitta$^2$}
\affiliation{%
$^1$Department of Physics, Kyoto University, Oiwake-cho, Kitashirakawa, Sakyo-ku, Kyoto 606-8502, Japan, \\
$^2$Department of Physics, and Research and
    Education Center for Natural Sciences, Keio University, Hiyoshi
    4-1-1, Yokohama, Kanagawa 223-8521, Japan
}%

\date{\today}

\begin{abstract}
We study Nambu-Goldstone (NG) modes 
or gapless modes
propagating along a skyrmion (lump) line 
in a relativistic and non-relativistic $O(3)$ sigma model,
the latter of which describes isotropic 
Heisenberg ferromagnets. 
We show for the non-relativistic case 
that there appear two coupled gapless modes 
with a quadratic dispersion. 
In addition to the well-known Kelvin mode  
consisting of two translational zero modes 
transverse to the skyrmion line, 
we show that the other consists of a 
magnon and dilaton, that is, 
a NG mode for 
a spontaneously broken 
internal $U(1)$ symmetry 
and a quasi-NG mode for  
a spontaneously broken scale 
symmetry of the equation of motion.
We find that 
the commutation relations of Noether charges admit 
a central extension between 
the dilatation and phase rotation,  
in addition to the one between two translations
found recently.
The counting rule is consistent with 
the Nielsen-Chadha and Watanabe-Brauner relations 
only when we take into account quasi-NG modes.

\end{abstract}

\pacs{05.30.Jp, 03.75.Lm, 03.75.Mn, 11.27.+d}

\maketitle

\section{Introduction}

When the symmetry of a Hamiltonian or Lagrangian 
is broken in the ground state, it is said that 
the symmetry is spontaneously broken.
When a continuous symmetry is spontaneously broken, 
there appear Nambu-Goldstone (NG) modes 
as gapless excitations that are dominant at low-energy. 
It is enough to incorporate these degrees of freedom 
to construct low-energy effective theories. 
In relativistic theories, there is a one-to-one correspondence 
between broken symmetry generators and NG modes, 
at least for internal symmetries.
In the non-relativistic cases, this is not so; 
In addition to type-I (relativistic) NG modes 
with a linear dispersion relation 
each of which corresponds to one broken generator,  
there are type-II (non-relativistic) NG modes
with a quadratic dispersion relation, 
each of which corresponds to two broken generators \cite{Nielsen:1975hm,Nambu:2004yia,
Watanabe:2011ec,Watanabe:2012hr,Hidaka:2012ym}. 
The criteria were summarized as 
the Watanabe-Brauner relation \cite{Watanabe:2011ec} 
stating that when 
two broken generators, $X_i$ and $X_j$, 
do not commute in the ground state, 
$\left<[X_i,X_j]\right> \neq 0$, 
they give rise to one type-II NG mode,  
such as magnons in ferromagnets. 
This has been recently proved  for internal symmetries \cite{Watanabe:2012hr,Hidaka:2012ym}.

However, there are no general arguments for 
spontaneously broken space-time symmetry 
(see, e.g., Refs.~\cite{Watanabe:2013iia,Hayata:2013vfa,Brauner:2014aha}
 for recent studies). 
In the presence of topological defects, 
space-time symmetries are spontaneously broken.  
For instance, it has been long known that 
there appears only one type-II NG mode, 
known as a Kelvin mode, or Kelvon if quantized, 
corresponding to two translational symmetries
spontaneously broken in the presence of 
a quantized vortex  in superfluids or a skyrmion (lump) 
\cite{Polyakov:1975yp} 
in ferromagnets,
while there are two type-I NG modes in 
the case of a relativistic string 
(see, e.g., Ref.~\cite{Kobayashi:2013gba,Eto:2013hoa} 
for a vortex and  
Ref.~\cite{skyrmion-dynamics} for a skyrmion).
Recently, Watanabe and Murayama \cite{Watanabe:2014pea}
have found 
\begin{align}
 \left[P_x, P_y\right] = B \neq 0 \label{eq:alg0}
\end{align}
in the background of a quantized vortex or 
a skyrmion line, where $P_x$ and $P_y$ are 
the Noether charges of translations 
perpendicular to the skyrmion  or vortex line, 
and $B$ is a topological charge for 
the skyrmion or vortex line. 
The two translational generators give 
one type-II NG mode, a Kelvon,
to be consistent with the 
 Watanabe-Brauner relation \cite{Watanabe:2011ec}.  
In our previous work 
\cite{Kobayashi:2014xua}, we have further found 
\begin{align}
 \left[ P_x, \Theta \right] = W \neq 0  \label{eq:alg}
\end{align}
in the background of a domain wall 
\cite{Abraham:1992vb} (a magnetic domain wall 
in ferromagnets \cite{Tatara:2008}),  
where $P_x$ is the Noether charge of 
the translation perpendicular to the wall, 
$\Theta$ is the Noether charge of an internal $U(1)$ symmetry, 
and $W$ is a topological charge of the domain wall.
A similar result has been obtained in Ref.~\cite{Watanabe:2014zza} 
for a domain wall in 
two-component Bose-Einstein condensates 
\cite{Takeuchi:2013mwa}. 
In the relativistic case, the two operators in Eq.~(\ref{eq:alg})
commute and there are two type-I NG modes.
The latter non-commutative relation (\ref{eq:alg}) resembles
supersymmetry algebras  
in the presence of Bogomol'nyi-Prasad-Sommerfield 
(BPS) solitons in supersymmetric field theories  \cite{Witten:1978mh,Dvali:1996bg}
and $p$-branes in supergravity and string theory 
\cite{de Azcarraga:1989gm}.

In this paper, 
in addition to Eq.~(\ref{eq:alg0}) for 
the translational modes in the presence of a skyrmion line, 
we show 
\begin{align}
 \left[D, \Theta + M_{12}\right] 
\neq 0
\end{align}
where $D$ is the Noether charge of 
a dilatation and 
$M_{12}$ is the Noether charge of 
a rotation around the $z$--axis 
along which the skyrmion line is placed. 
The skyrmion solution has four moduli 
$X$, $Y$, $\theta$, and $R$. 
While $X$, $Y$ and $\theta$ are NG modes
of two translations and the internal $U(1)$ symmetry, 
respectively, 
we point out that a dilaton $R$ is 
a quasi-NG mode \cite{Weinberg:1972fn}, 
which appears when a symmetry of the equation of 
motion, but not that of the Lagrangian or action,  
is spontaneously broken. 
By constructing the low-energy effective field theory 
on a 1+1 dimensional skyrmion world-sheet via 
the moduli approximation \cite{Manton:1981mp},
we find that 
the dilaton $R$ and the $U(1)$ NG mode (magnon) $\theta$ are coupled 
to give rise to one type-II gapless mode,
which is consistent with the Watanabe-Brauner relation 
only when we count quasi-NG modes, 
while, in the relativistic case, 
the dilaton and the $U(1)$ NG mode appear independently 
as type-I (quasi-)NG modes. 
We further study fluctuations around the solution 
in the Bogoliubov analysis and find the same result. 
We also consider the effects of explicit breaking 
terms for the scale symmetry, 
that is the baby skyrme term and a potential term.
In the presence of these terms, 
the skyrmions are known as baby skyrmions
\cite{Piette:1994ug}.
We show that these terms introduce 
a potential term for the dilaton 
 in the effective Lagrangian.
In the relativistic case, the dilaton becomes massive 
and the magnon remains massless as expected.
In the non-relativistic case, on the other hand, 
the magnon-dilaton becomes a type-I NG mode.

\section{Nonlinear sigma models and skyrmions}
We consider the following relativistic and non-relativistic $\mathbb{C}P^1$ Lagrangian densities $\mathcal{L}\sub{rel}$ and $\mathcal{L}\sub{nrel}$,
\begin{align}
\mathcal{L}\sub{rel} = \frac{|\dot{u}|^2 - |\nabla u|^2}{(1 + |u|^2)^2}, \quad
\mathcal{L}\sub{nrel} = \frac{i ( u^\ast \dot{u} - \dot{u}^\ast u)}{2 (1 + |u|^2)} - \frac{|\nabla u|^2}{(1 + |u|^2)^2}, \label{eq-Lagrangian}
\end{align}
where, $u \in \mathbb{C}$ is the complex projective coordinate of $\mathbb{C}P^1$, 
defined as $\phi^T = (1, u)^T / \sqrt{1 + |u|^2}$ with 
normalized two complex 
scalar fields $\phi = (\phi_1, \phi_2)^T$.
The non-relativistic Lagrangian $\mathcal{L}\sub{nrel}$ is obtained by taking the non-relativistic limit of $\mathcal{L}\sub{rel}$ (see Appendix A of Ref. \cite{Kobayashi:2014xua}).
$\mathcal{L}\sub{rel}$ and $\mathcal{L}\sub{nrel}$ 
can be rewritten as $O(3)$ nonlinear sigma models,
\begin{align}
\begin{split}
\mathcal{L}\sub{rel} = \frac{1}{4} \{ |\dot{\Vec{n}}|^2 - |\nabla \Vec{n}|^2 \}, \quad
\mathcal{L}\sub{nrel} = \frac{\dot{n}_1 n_2 - n_1 \dot{n}_2}{2 (1 + n_3)} - \frac{1}{4} |\nabla \Vec{n}|^2, \label{eq-sigma}
\end{split}
\end{align}
under the Hopf map for a three-vector of real scalar fields $\Vec{n} \equiv \phi^\dagger  \Vec{\sigma}  \phi $ with the Pauli matrices $\Vec{\sigma}$.
These models describe isotropic Heisenberg ferromagnets. 
In this paper, we use the $\mathbb{C}P^1$ model notation.

The Lagrangians $L\sub{rel} = \int d^3x\: \mathcal{L}\sub{rel}$ and $L\sub{nrel} = \int d^3x\: \mathcal{L}\sub{nrel}$ are invariant under a global $SU(2)$ rotation of $\phi$, the Poincar\'e (for $L\sub{rel}$) or Galilean (for $L\sub{nrel}$) transformation. 
In the vacuum of the system, {\it i.e.}, the arbitrary uniform $u$, the internal $SU(2)$ symmetry is spontaneously broken down to a $U(1)$ symmetry with the identification of the global phase of $\phi$ (phase of $\phi_1$).
The vacuum manifold is, therefore, isomorphic to $\mathcal{M}_1 \simeq SU(2) / U(1) \simeq \mathbb{C}P^1 \simeq S^2$.

The dynamics of $u$ can be described by the Euler-Lagrange equation for $\mathcal{L}\sub{rel}$ and $\mathcal{L}\sub{nrel}$:
\begin{align}
\begin{split}
& (1 + |u|^2) \ddot{u} - 2 u^\ast \dot{u}^2 \stackrel{\mathrm{rel}}{=} (1 + |u|^2) \nabla^2 u - 2 u^\ast (\nabla u)^2, \\
& - i (1 + |u|^2) \dot{u} \stackrel{\mathrm{nrel}}{=} (1 + |u|^2) \nabla^2 u - 2 u^\ast (\nabla u)^2, \label{eq-dynamics}
\end{split}
\end{align}
where $\stackrel{\mathrm{rel}}{=}$ and $\stackrel{\mathrm{nrel}}{=}$ correspond to dynamics derived from $\mathcal{L}\sub{rel}$ and $\mathcal{L}\sub{nrel}$, respectively. 
The equations of motion (\ref{eq-dynamics}) enjoy 
an additional symmetry of a scaling transformation 
$(t,x,y,z) \to (st,sx,sy,sz)$ for the relativistic case and $(t,x,y,z) \to (s^2t,sx,sy,sz)$ for the non-relativistic case 
with $s \in {\mathbb R}^+ -\{0\} \simeq {\mathbb R}$.
These are not symmetry of the Lagrangians or the actions.

We next consider a static skyrmion line solution.
A straight skyrmion line solution parallel to the $z$-axis is
\cite{Polyakov:1975yp}
\begin{align}
u_0(x, y, z) = \frac{\exp \big\{ i \big( \tan^{-1}\frac{y - Y}{x - X} + \theta \big) \big\} \sqrt{(x - X)^2 + (y - Y)^2} }{R_0 + R}, \label{eq-skyrmion}
\end{align}
where $R_0 \in \mathbb{R}$ is the characteristic radius of the skyrmion line, and $X,Y \in \mathbb{R}$, $\theta\ (0 \leq \theta < 2 \pi)$, and $R \in \mathbb{R}$ are the translational, phase, and dilatation moduli of the skyrmion line respectively.
The tension of the skyrmion line (the energy per unit area) is
\begin{align}
T = \int dx dy \: \frac{|\nabla u_0|^2}{(1 + |u_0|^2)^2} = 2 \pi,
\end{align}
independent of $X$, $Y$, $\theta$, and $R$.
The $U(1)$  phase rotation of $u$, 
the translation $\mathbb{R}^2_{txy}$ inside the $xy$--plane, and  the dilatation $\mathbb{R}_{dxy}$ inside 
the $xy$--plane 
are spontaneously broken 
in the vicinity of the skyrmion line.
The four moduli $X$, $Y$, $\theta$, and $R$ in Eq. \eqref{eq-skyrmion} are regarded as NG modes corresponding to $\mathbb{R}^2_{txy}$, $U(1)$ and $\mathbb{R}_{dxy}$, respectively, 
localized in the vicinity of the skyrmion line \cite{rotational-symmetry}. 
The NG modes $X$ and $Y$ are the translational modes,
which are called as Kelvin waves of the skyrmion line 
in the non-relativistic case.
The NG mode $\theta$ 
may be called as the localized magnon. 
We call $R$ as the dilaton, 
associated with
the spontaneously broken $\mathbb{R}_{dxy}$. 
This $\mathbb{R}_{dxy}$ is merely a symmetry of 
the equation of motion but not that of 
the full theory. 
Consequently $R$ is a so-called quasi-NG 
mode 
\cite{Weinberg:1972fn} 
but not a genuine NG mode, 
which is gapless at least classically,  
see, e.g., Ref.~\cite{Uchino:2010pf} for 
an example of a quasi-NG mode in 
condensed matter physics.
The dilaton $R$ is similar to the localized varicose mode excited along a superfluid vortex in terms of the radius wave of the string \cite{Simula:2008},
but it has a gap 
in the absence of the dilatational symmetry.

\section{Low-energy effective theory of a skyrmion line}
We next discuss the dynamics of the localized NG modes in the vicinity of the  skyrmion line, 
by constructing the effective theory on a skyrmion line  
using the moduli approximation \cite{Manton:1981mp}.
Let us introduce the $z$ and $t$ dependences of the moduli in Eq.~\eqref{eq-skyrmion} as $X(z,t)$, $Y(z,t)$, $\theta(z,t)$, and $R(z,t)$:
\begin{align}
u(x, y, z) = \frac{\exp \big[ i \big\{ \tan^{-1}\frac{y - Y(z,t)}{x - X(z,t)} + \theta(z,t) \big\} \big] \sqrt{\{x - X(z,t)\}^2 + \{y - Y(z,t)\}^2} }{R_0 + R(z,t)}. \label{eq-effective-theory-ansatz}
\end{align}
By inserting Eq.~\eqref{eq-effective-theory-ansatz} back into Eq.~\eqref{eq-Lagrangian}, 
the two effective Lagrangians $L\up{eff}\sub{rel}$ and $L\up{eff}\sub{nrel}$ defined as $L\up{eff}\sub{rel} = 
\int_{-L}^L dx\: \int_{-\sqrt{L^2 - x^2}}^{\sqrt{L^2 - x^2}} dy\: \mathcal{L}\sub{rel}$ and $L\up{eff}\sub{nrel} = 
\int_{-L}^L dx\: \int_{-\sqrt{L^2 - x^2}}^{\sqrt{L^2 - x^2}} dy\: \mathcal{L}\sub{nrel}$ can be calculated, to yield 
\begin{align}
\begin{split}
& L\up{eff}\sub{rel} = \pi (\dot{X}^2 + \dot{Y}^2 - X_z^2 - Y_z^2) + 2 \pi \log\bigg(\frac{L}{R_0}\bigg) ( R_0^2 \dot{\theta}^2 + \dot{R}^2 - R_0^2 \theta_z^2 - R_z^2 ) \\
&\phantom{\frac{L\up{eff}\sub{rel}}{\pi} =} - 2 \pi + O(\partial_z^3), \\
& L\up{eff}\sub{nrel} = - \pi L^2 \dot{\theta} + \pi (\dot{X} Y - X \dot{Y} - X_z^2 - Y_z^2) + 2 \pi \log\bigg(\frac{L}{R_0}\bigg) (2 R_0 \dot{\theta} R - R_0^2 \theta_z^2 - R_z^2) \\
&\phantom{\frac{L\up{eff}\sub{nrel}}{\pi} =} - 2 \pi + O(\partial_z^3),
\end{split}
\end{align}
up to the quadratic order in derivatives.

The low-energy dynamics of $X$, $Y$, $\theta$, and $R$ derived from the Euler-Lagrange equation becomes
\begin{subequations}
\begin{align}
& \ddot{X} \stackrel{\mathrm{rel}}{=} X_{zz}, \quad
\ddot{Y} \stackrel{\mathrm{rel}}{=} Y_{zz}, \quad
\ddot{\theta} \stackrel{\mathrm{rel}}{=} \theta_{zz}, \quad
\ddot{R} \stackrel{\mathrm{rel}}{=} R_{zz}, \label{eq-skyrmion-dynamics-rel} \\
& \dot{X} \stackrel{\mathrm{nrel}}{=} - Y_{zz} , \quad
\dot{Y} \stackrel{\mathrm{nrel}}{=} X_{zz}, \quad
R_0 \dot{\theta} \stackrel{\mathrm{nrel}}{=} - R_{zz}, \quad
\dot{R} \stackrel{\mathrm{nrel}}{=} R_0 \theta_{zz} .
  \label{eq-skyrmion-dynamics-nrel}
\end{align}
\end{subequations}
For the relativistic case, all dynamics of $X$, $Y$, $\theta$ and $R$ are independent of each other, giving 
linear dispersions:
\begin{align}
\omega\sub{rel} = \pm |\Vec{k}|, \label{eq-dispersion-rel}
\end{align}
with the frequencies $\omega\sub{rel}$, and the wave number $k$.
Oscillations $X$ and $Y$ of the skyrmion line 
into the $x$ and $y$--directions,
a localized magnon $\theta$ and 
a dilaton $R$ independently 
propagate along the $z$-axis.

\begin{figure}[tbh]
\centering
\includegraphics[width=0.7\linewidth]{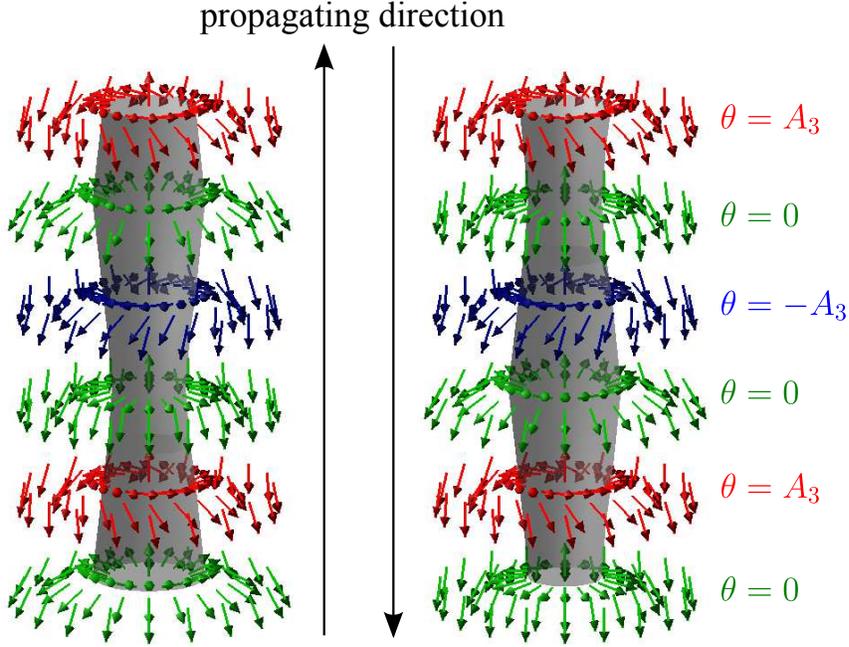}
\caption{\label{fig-dilatation} (Color online) Schematic picture of the localized coupled  magnon-dilaton for 
Eq.~\eqref{eq-dilatation}.
The arrows show the direction of $\Vec{n} = \phi^\dagger \Vec{\sigma} \phi$ with $\phi^T = (1,u)^T / \sqrt{1 + |u|^2}$ and their colors shows the value of $\theta$.
The transparent surface shows the isosurface for $|u| = 0$ ($n_3 = 0$).
For left (right) figures, the coupled localized magnon-dilaton propagate in the upper (lower) direction.
}
\end{figure}

Being different from the relativistic case, there are two different coupled modes, $X$ and $Y$, and $\theta$ and $R$ in the non-relativistic case.
Typical solutions of Eq.~\eqref{eq-skyrmion-dynamics-nrel} are
\begin{subequations}
\begin{align}
& X = A_{(XY)\pm} \sin(k z \mp \omega\sub{nrel} t + \delta_{(XY)\pm}), \quad
Y = \mp A_{(XY)\pm} \cos(k z \mp \omega\sub{nrel} t + \delta_{(XY)\pm}), \label{eq-kelvin} \\
& \theta = A_{(\theta R)\pm} \sin(k z \mp \omega\sub{nrel} t + \delta_{(\theta R)\pm}), \quad
R = \mp R_0 A_{(\theta R)\pm} \cos(k z \mp \omega\sub{nrel} t + \delta_{(\theta R)\pm}), \label{eq-dilatation}
\end{align}
\end{subequations}
where $A_{(XY),(\theta R)\pm} \in \mathbb{R}$ and $\delta_{(XY),(\theta R)\pm} \in \mathbb{R}$ are arbitrary constants.
Waves of $X$ and $Y$ couple and 
propagate as a spiral Kelvin wave, 
and $\theta$ and $R$ couple to each other and 
propagate as a coupled magnon-dilaton, 
both with a quadratic dispersion
\begin{align}
\omega\sub{nrel} = k^2. \label{eq-dispersion-nrel}
\end{align}
For the upper and lower signs in Eqs.~\eqref{eq-kelvin} and \eqref{eq-dilatation}, each coupled NG mode propagates in the direction parallel and anti-parallel to $z$--axis, respectively.
In contrast to the Kelvin wave in Eq.~\eqref{eq-kelvin} which are combinations of two translational modes in real space, the localized coupled magnon-dilaton mode in Eq. \eqref{eq-dilatation} is a combination of the phase mode of the internal degrees of freedom and the dilatation in real space.
Figure \ref{fig-dilatation} shows the schematic picture of coupled localized magnon-dilaton mode for Eq. \eqref{eq-dilatation} 
\cite{movie}. 

\section{Linear response theory}
Linear response theory is another technique to study the dynamics of the gapless modes $X,Y,\theta,R$.
Let us consider the ansatz of the straight skyrmion line solution and its fluctuation: $u = u_0 + \delta u = u_0 + a_+ e^{i (k z - \omega t)} + a_-^\ast e^{- i (k z - \omega t)}$.
By inserting this ansatz into the dynamical equation \eqref{eq-dynamics}, 
the Bogoliubov-de Gennes equations are obtained, 
\begin{align}
\begin{split}
& \omega\sub{rel}^2 a_{\pm} \stackrel{\mathrm{rel}}{=} \bigg\{ (k^2 - \nabla_{\Vec{r}}^2) + \frac{4 (r \partial_r \pm i \partial_\theta)}{r^2 + R_0^2} \bigg\} a_{\pm} + O(a_{\pm}^2), \\
& \omega_{\mathrm{nrel},\pm} a_{\pm} \stackrel{\mathrm{nrel}}{=} \pm \bigg\{ (k^2 - \nabla_{\Vec{r}}^2) + \frac{4 (r \partial_r \pm i \partial_\theta)}{r^2 + R_0^2} \bigg\} a_{\pm} + O(a_{\pm}^2), \label{eq-BdG}
\end{split}
\end{align}
up to the linear order of $a_\pm$.
Here, $\nabla_{\Vec{r}} = (\partial_x, \partial_y)$ denotes 
the derivative in the $xy$--plane.
By expanding $a_{\pm}$ as $a_{\pm} = \sum_l a_{\pm,l} e^{i l \phi}$, we obtain
\begin{align}
\begin{split}
& \omega\sub{rel}^2 a_{\pm,l} \stackrel{\mathrm{rel}}{=} \bigg\{ (k^2 - \partial_r^2 - \partial_r / r + l^2 / r^2 ) + \frac{4 (r \partial_r \mp l)}{r^2 + R_0^2} \bigg\} a_{\pm,l} + O(a_{\pm}^2), \\
& \omega_{\mathrm{nrel},\pm} a_{\pm,l} \stackrel{\mathrm{nrel}}{=} \pm \bigg\{ (k^2 - \partial_r^2 - \partial_r / r + l^2 / r^2) + \frac{4 (r \partial_r \mp l)}{r^2 + R_0^2} \bigg\} a_{\pm,l} + O(a_{\pm}^2).
\end{split}
\end{align}
There are two characteristic solutions related to the Kelvin wave, localized magnon, and dilaton: $a_{\pm,0} = 1$ and $a_{\pm,1} = r / R_0$, and eigenvalues are $\omega\sub{rel}^2 = k^2$ and $\omega\sub{nrel,\pm} = \pm k^2$.
For the relativistic case, NG modes for $X$, $Y$, $\theta$, and $R$ are obtained as
\begin{align}
\begin{split}
&X\ :\ \delta u = A_{X\pm} (a_{+,0} e^{i (k z \mp k t + \delta_{X\pm})} + a_{-,0} e^{- i (k z \mp k t + \delta_{X\pm})}), \\
&Y\ :\ \delta u = A_{Y\pm} (a_{+,0} e^{i (k z \mp k t + \delta_{Y\pm})} - a_{-,0} e^{- i (k z \mp k t + \delta_{Y\pm})}), \\
&\theta\ :\ \delta u = A_{\theta\pm} (a_{+,1} e^{i \phi} e^{i (k z \mp k t + \delta_{\theta\pm})} - a_{-,1} e^{i \phi} e^{- i (k z \mp k t + \delta_{\theta\pm})}), \\
&R\ :\ \delta u = A_{R\pm} (a_{+,1} e^{i \phi} e^{i (k z \mp k t + \delta_{R\pm})} + a_{-,1} e^{i \phi} e^{- i (k z \mp k t + \delta_{R\pm})}),
\end{split} \label{eq-NG-mode-rel}
\end{align}
with arbitrary constant $A_{X,Y,\theta,R\pm}, \delta_{X,Y,\theta,R\pm} \in \mathbb{R}$.
The upper (lower) sign in Eq.~\eqref{eq-NG-mode-rel} shows the NG modes propagating in the direction parallel (anti-parallel) to the $z$--axis.
For non-relativistic case, the coupled NG modes for, $(X,Y)$, and $(\theta,R)$ are obtained as
\begin{align}
\begin{split}
&(X,Y)\ :\ \delta u = A_{(XY)\pm} a_{\pm,0} e^{\pm i (k z \mp k^2 t + \delta_{(XY)\pm})}, \\
&(\theta,R)\ :\ \delta u = A_{(\theta R)\pm} a_{\pm,1} e^{i \phi} e^{\pm i (k z \mp k^2 t + \delta_{(\theta R)\pm})}.
\end{split} \label{eq-NG-mode-nrel}
\end{align}

We shortly note that there are countably infinite number of gapless solutions to the Bogoliubov-de Gennes equation \eqref{eq-BdG}, $a_{\pm,n} = r^n / R_0^n$ ($n \in \mathbb{Z}_0^+)$ and corresponding zero modes $\omega\sub{rel}^2 = k^2$ and $\omega\in{nrel,\pm} = \pm k^2$ besides the present solutions $a_{\pm,0}$ and $a_{\pm,1}$.
Solutions for $n = 0,1,2$ correspond to 
the (quasi-)NG modes; {\it i.e.}, $a_{\pm,0}$ corresponds to the Kelvin waves,  
$a_{\pm,1}$ corresponds to 
the localized magnon and dilaton,
and $a_{\pm,2}$ corresponds to the bulk magnon far from the skyrmion line,
for which we have not discussed in this paper.
The other solutions, $a_{\pm,n\neq 0,1,2}$, do not originate from any symmetry of the Lagrangians and 
cannot be regarded as  
NG modes.
We will soon discuss this in detail elsewhere.

\section{Commutation relation}

By the two techniques of the effective theory with the moduli approximation and the linear-response theory, 
we have shown the independence of the four gapless modes in the presence of the skyrmion line-- the two translational modes, the localized magnon, and the dilaton.
They are independent of each other with the linear dispersion relations \eqref{eq-dispersion-rel} for the relativistic theory with $\mathcal{L}\sub{rel}$, while the coupled spiral Kelvin wave and the coupled localized magnon-dilaton are formed showing the quadratic dispersion relations \eqref{eq-dispersion-nrel} for the non-relativistic theory with $\mathcal{L}\sub{nrel}$.
These modes are (quasi-)NG modes appearing 
as a consequence of the spontaneous breaking 
continuous symmetries; 
the $\mathbb{R}_{txy}^2$ translational symmetry for Kelvin waves, the $U(1)$ symmetry for the localized magnon, and the two-dimensional $\mathbb{R}_{dxy}$ scaling symmetry for the dilaton.
The Lorentz invariance in the relativistic model supports that the number of (quasi-)NG modes is equivalent to that of symmetry generators $N\sub{BG}$ corresponding to spontaneously broken symmetries, and all (quasi-)NG modes are type-I for the linear dispersion \eqref{eq-dispersion-rel}.
Without the Lorentz invariance, there appear not only type I (quasi-)NG modes but also type II (quasi-)NG modes with the quadratic dispersion and the relation between the number of (quasi-)NG modes and $N\sub{BG}$ becomes more complicated.
In both cases, the numbers of (quasi-)NG modes saturates the equality of the Nielsen-Chadha inequality \cite{Nielsen:1975hm}, $N\sub{I} + 2 N\sub{II} \geq N\sub{BG}$, where $N\sub{I}$ and  $N\sub{II}$ are the total numbers of the type-I NG modes and the type-II NG modes.
In the case of internal symmetries,
it has been shown 
in Refs.~\cite{Watanabe:2012hr,Hidaka:2012ym} 
that the equality of the Nielsen-Chadha inequality is saturated as the Watanabe-Brauner's relation \cite{Watanabe:2011ec},
\begin{align}
N\sub{BG} - N\sub{NG} = \frac{1}{2} \mathrm{rank} \rho, \quad
\rho_{i,j} = \lim_{V \to \infty} \frac{1}{V} \int d^3\Vec{x} [\Omega_i, \Omega_j] 
 \Big|_{u = u_0}. \label{eq-Watanabe-Brauner}
\end{align}
Here, $N\sub{NG} = N\sub{I} + N\sub{II}$ is the total number of NG modes, $V$ is the volume of the system, 
$\Omega_i$ is the Noether charge or a generator of broken symmetries, 
and $[\cdot,\cdot]$ is 
a commutator or the Poisson bracket in the classical level.
We see that a mismatching $N\sub{BG} \neq N\sub{NG}$ takes place when 
commutators of  
broken generators are non-vanishing in the ground states  
in non-relativistic theories.
This relation has been proven for cases of the bulk magnons (two internal symmetries) and the Kelvin wave (two space-time symmetries) in the massless $\mathbb{C}P^1$ model for the isotropic Heisenberg ferromagnet \cite{Watanabe:2014pea}.
It has been also proved for 
the translational and internal $U(1)$ zero modes
in the background of a domain wall 
in the massive $\mathbb{C}P^1$ model for the Heisenberg ferromagnet with one easy axis \cite{Kobayashi:2014xua}. 
As the case of a domain wall in Ref.~\cite{Kobayashi:2014xua}, 
the two broken generators corresponding to the coupled localized magnon-dilaton, that is,
the internal $U(1)$ symmetry and translational symmetry,  
intuitively commute, because underlying symmetries are 
the direct product and are independent of each other, {\it i.e.}, $U(1) \times \mathbb{R}_{dxy}$. 
In order to check whether the relation \eqref{eq-Watanabe-Brauner} also holds in our case or not, let us directly calculate the commutation relation between 
Noether charges of symmetry generatros of the localized magnon and the dilaton. 

Before calculating the commutator for the localized magnon and the dilatation mode, we briefly overview the commutator of the two translations for the spiral Kelvin wave \cite{Watanabe:2014pea} which also intuitively commute with each other.
Let us define the momenta $v$ conjugate to $u$ as
\begin{align}
v \stackrel{\mathrm{rel}}{=} \frac{\partial \mathcal{L}\sub{rel}}{\partial \dot{u}} = \frac{\dot{u}^\ast}{(1 + |u|^2)^2}, \quad
v \stackrel{\mathrm{nrel}}{=} \frac{\partial \mathcal{L}\sub{nrel}}{\partial \dot{u}} = \frac{i u^\ast}{2 (1 + |u|^2)}.
\end{align}
Then, the Noether's charges for the translations for $x$ and $y$--directions are obtained as 
\begin{align}
& P_x = \int d^2x\: J^0_X , \quad J^0_X = u_x v, \qquad 
P_y = \int d^2x\: J^0_Y , \quad J^0_Y = u_x v.
\end{align}
The commutator between $P_x$ and $P_y$
can be calculated from 
$[u(x_1,y_1), v(x_2,y_2)] = \delta(x_1 - x_2) \delta(y_1 - y_2) \equiv \delta^2(\Vec{x}_1 - \Vec{x}_2)$, to yield 
\begin{align}
\begin{split}
[P_x, P_y] &= \int d^2x_1\: \int d^2x_2\: [J^0_X(x_1,y_1), J^0_Y(x_2,y_2)] \\
&= \int d^2x_1\: \int d^2x_2\: [u_{x_1}(x_1,y_1) v(x_1,y_1), u_{y_2}(x_2,y_2) v(x_2,y_2)] \\
&= \int d^2x_1\: \int d^2x_2\: \{ u_{x_1}(x_1,y_1) [ v(x_1,y_1), u_{y_2}(x_2,y_2)] v(x_2,y_2) \\
&\phantom{= i \int d^2x_1\: \int d^2x_2\: \{} + u_{y_2}(x_2,y_2) [u_{x_1}(x_1,y_1), v(x_2,y_2)] v(x_1,y_1) \} \\
&= \int d^2x_1\: \int d^2x_2\: \{ u_{x_1}(x_1,y_1) v_{y_2}(x_2,y_2) - u_{y_2}(x_2,y_2) v(x_1,y_1) \} \delta^2(\Vec{x}_1 - \Vec{x}_2) \\
&= \int d^2x\: \{ u_{x}(x,y) v_{y}(x,y) - u_{y}(x,y) v(x,y) \} \\
&= \int d^2x\: \frac{u_r v_\phi - u_\phi v_r}{r},
\end{split}
\end{align}
with the cylinderical coordinate $(r, \phi)$.

The commutator vanishes for the relativistic case
because of $v=0$ ($\dot u=0$), implying two type I NG modes.

For the non-relativistic case, 
the commutator becomes \cite{Watanabe:2014pea}
\begin{align}
\begin{split}
[P_x, P_y]  = \int d^2x\: \frac{u_r v_\phi - u_\phi v_r}{r} = \int d^2x\: b = B \neq 0, \label{eq:Px-Py}
\end{split}
\end{align}
where $b$ and $B$ are the topological charge density and the total topological charge of the skyrmion line:
\begin{align}
b = u_x v_y - u_y v_x = \frac{u_r v_\phi - u_\phi v_r}{r} = \frac{R_0^2}{(r^2 + R_0^2)^2}, \quad
B = \int_0^\infty dr\: \int_0^{2 \pi} d\theta\: r b = \pi \label{eq:b}.
\end{align}
In the last equalities in 
\eqref{eq:b}, we have used 
the background of a single skyrmion line 
$u = u_0 = r e^{i \phi} / R_0$.

We next calculate the commutator for the localized magnon and dilaton.
The Noether's charges for the phase shift and the dilatation are obtained as 
\begin{align}
& \Theta =  \int d^2x\: J^0_\theta , \quad J^0_\theta = i u v, \\ 
& D = \int d^2x\: J^0_R , \quad J^0_R = (x u_x + y u_y) v,
\end{align}
respectively. 
The commutator between them reads
\begin{align}
\begin{split}
[D, \Theta] &= \int d^2x_1\: \int d^2x_2\: [J^0_D(x_1,y_1), J^0_\theta(x_2,y_2)] \\
&= i \int d^2x_1\: \int d^2x_2\: [\{x_1 u_{x_1}(x_1,y_1) + y_1 u_{y_1}(x_1,y_1)\} v(x_1,y_1), u(x_2,y_2) v(x_2,y_2)] \\
&= i \int d^2x_1\: \int d^2x_2\: [ x_2 \{ u(x_1,y_1) [ u_{x_2}(x_2,y_2), v(x_1,y_1)] v(x_2,y_2) \\
&\phantom{= i \int d^2x_1\: \int d^2x_2\: [ x_2 \{} + u_{x_2}(x_2,y_2) [v(x_2,y_2), u(x_1,y_1)] v(x_1,y_1) \} \\
&\phantom{= i \int d^2x_1\: \int d^2x_2\: [ } + y_2 \{ u(x_1,y_1) [ u_{y_2}(x_2,y_2), v(x_1,y_1)] v(x_2,y_2) \\
&\phantom{= i \int d^2x_1\: \int d^2x_2\: [ + y_2 \{} + u_{y_2}(x_2,y_2) [v(x_2,y_2), u(x_1,y_1)] v(x_1,y_1) \}]. \\
&= - i \int d^2x_1\: \int d^2x_2\: [ x_2 \{ u(x_1,y_1) v_{x_2}(x_2,y_2) + u_{x_2}(x_2,y_2) v(x_1,y_1) \} \\
&\phantom{= - i \int d^2x_1\: \int d^2x_2\: [} + y_2 \{u(x_1,y_1) v_{y_2}(x_2,y_2) + u_{y_2}(x_2,y_2) v(x_1,y_1) \} \\
&\phantom{= - i \int d^2x_1\: \int d^2x_2\: [} + 2 u(x_1,y_1) v(x_2,y_2)] \delta^2(\Vec{x}_1 - \Vec{x}_2) \\
&= - i \int d^2x\: [ r \{ u_r(r,\phi) v(r,\phi) + u(r,\phi) v_r(r,\phi) \} + 2 u(r,\phi) v(r,\phi)]. \label{eq-D-Theta-detail} 
\end{split}
\end{align}

For the non-relativistic case, 
the commutator becomes
\begin{align}
\begin{split}
[D, \Theta]  = \int d^2x\: \frac{r^2 (r^2 + 2 R_0^2)}{(r^2 + R_0^2)^2} = \int d^2x\: r^2 \bigg( b + \frac{1}{r^2 + R_0^2} \bigg) \neq 0, \label{eq:D-Theta}
\end{split}
\end{align}
while it vanishes for the relativistic case. 

The ansatz in Eq. \eqref{eq-effective-theory-ansatz} with $X = Y = 0$, 
\begin{align}
u(r, \phi, z) = \frac{r \exp \{ i (\phi + \theta) \}}{R_0 + R},
\end{align}
implies that the localized magnon $\theta$ can be induced not only by the phase shift of $u$, $u \to u e^{i \theta}$, but also by a spatial rotation along $z$--axis, $\phi \to \phi + \theta$.
Therefore, we  further calculate the commutator between the spatial rotation and the dilatation.
The Noether's charge for the rotation is
\begin{align}
M_{12} = \int d^2x\: J_\phi^0, \quad J_\phi^0 = (x u_y - y u_x) v.
\end{align}
In addition to Eq. \eqref{eq-D-Theta-detail}, the commutator becomes
\begin{align}
\begin{split}
[D, M_{12}] &= \int d^2x_1\: \int d^2x_2\: [J^0_D(x_1,y_1), J^0_\phi(x_2,y_2)] \\
&= \int d^2x\: [ (x^2 + y^2) \{ u_x(x_2,y_2) v_y(x_1,y_1) - u_y(x_1,y_1) v_x(x_2,y_2) \} \\
&\phantom{= \int d^2x\: [} + 2 \{ y u_x(x_1,y_1) - x u_y(x_2,y_2) \} v(x,y) \} \\
&= \int d^2x\: [ r \{ u_r(r,\phi) v_\phi(r,\phi) - u_\phi(r,\phi) v_r(r,\phi) \} - 2 u_\phi(r,\phi) v(r,\phi)] \\
&= \int d^2x\: r^2 \bigg( b + \frac{1}{r^2 + R_0^2} \bigg)\\
& = [D, \Theta]. \label{eq-D-M-detaill}
\end{split}
\end{align}
The fact $[D, \Theta - M_{12}] = 0$ implies that $u_0$ is invariant under a simultaneous action of the phase shift $u \to u e^{i \theta}$ and the spatial rotation $\phi \to \phi - \theta$.
Therefore, we find an independent non-vanishing 
commutation relation 
\begin{align}
 \left[D,\Theta +M_{12} \right] 
= 2 \int d^2x\: r^2 \bigg( b + \frac{1}{r^2 + R_0^2} \bigg) 
\neq 0,
\end{align}
which is consistent with our result for the coupled localized 
magnon-dilaton.

\section{The explicit breaking term for the scale symmetry: 
the case of baby skyrmions}\label{sec:baby}

Here, we briefly investigate the effect of a small explicit breaking term for the scaling symmetry. 
One of the simple additional terms, $\mathcal{L}\sub{add}$, 
that explicitly breaks the scaling symmetry is
\begin{align}
\mathcal{L}\sub{add} = - \frac{\kappa \{ (|\nabla u|^2)^2 - |(\nabla u)^2|^2 \}}{(1 + |u|^2)^4} - \frac{2 \beta^2}{1 + |u|^2}.
\label{eq:baby}
\end{align}
Here, the first and second terms are the baby skyrme term and the potential term 
(corresponding to the magnetic field along 
the $n_3$ axis in ferromagnets),
by which the skyrmion tends to expand and shrink, 
respectively. 
In the presence of  both terms, the skyrmions 
are known as baby skyrmions with a fixed size 
\cite{Piette:1994ug}. 

Here, we suppose that the parameters 
$\kappa$ and $\beta$  are small, 
and treat the 
additional terms in Eq.~(\ref{eq:baby}) as small 
perturbations. 
We can assume the configuration in 
Eq. \eqref{eq-effective-theory-ansatz} 
is unchanged at the leading order. 
Minimizing the energy 
for the configuration in 
Eq. \eqref{eq-effective-theory-ansatz},  
the size $R_0$ is determined by 
\begin{equation}
\kappa = 3 \beta^2 R_0^4 \log(L / R_0).
\end{equation}  
With small $\kappa$ and $\beta$, the effective Lagrangians for 
$R$ and $\theta$ become
\begin{align}
\begin{split}
&L\sub{rel}\up{eff} + \int dx dy\: \mathcal{L}\sub{add} = 2 \pi \log\bigg(\frac{L}{R_0}\bigg) ( R_0^2 \dot{\theta}^2 + \dot{R}^2 - R_0^2 \theta_z^2 - R_z^2 - m R^2 ) + \mathrm{const} \\
&\phantom{L\sub{rel}\up{eff} + \int dx dy\: \mathcal{L}\sub{add} =} + O(\partial_z^3, R^3), \\
&L\sub{nrel}\up{eff} + \int dx dy\: \mathcal{L}\sub{add} = 2 \pi \log\bigg(\frac{L}{R_0}\bigg) ( 2 R_0 R \dot{\theta} - R_0^2 \theta_z^2 - R_z^2 - m R^2 ) + \mathrm{const} + O(\partial_z^3, R^3),
\end{split}
\end{align}
up to the second order in $R$, 
where the mass $m$ of $R$ is given by $m = 8 \beta^2$.
The effective Lagrangians for 
$X$ and $Y$ are unchanged.

The solutions of the Euler-Lagrange equation become
\begin{subequations}
\begin{align}
& \theta \stackrel{\mathrm{rel}}{=} A_{(\theta) \mathrm{rel}} \sin(k z - \omega_{(\theta) \mathrm{rel}} t + \delta_{(\theta) \mathrm{rel}}), \quad
R \stackrel{\mathrm{rel}}{=} A_{(R) \mathrm{rel}} \sin(k z - \omega_{(R) \mathrm{rel}} t + \delta_{(R) \mathrm{rel}}), \\
& \theta \stackrel{\mathrm{nrel}}{=} A\sub{nrel} \cos(k z - \omega\sub{nrel} t + \delta\sub{nrel}), \quad
R  \stackrel{\mathrm{nrel}}{=} \frac{A\sub{nrel} k}{\sqrt{m + k^2}} \sin(k z - \omega\sub{nrel} t + \delta\sub{nrel}),
\end{align}
\end{subequations}
with dispersions,
\begin{align}
\omega_{(\theta) \mathrm{rel}} = \pm |\Vec{k}|,\quad
\omega_{(R) \mathrm{rel}} = \pm \sqrt{\Vec{k}^2 + m},\quad
\omega\sub{nrel} = \frac{m |\Vec{k}| + |\Vec{k}|^3}{\sqrt{m + \Vec{k}^2}} = \sqrt{m} |\Vec{k}| + O(|\Vec{k}|^3),
\end{align}
where, $A_{(\theta,R) \mathrm{rel}},\ A\sub{nrel},\ \delta_{(\theta,R) \mathrm{rel}},\ \delta\sub{nrel} \in \mathbb{R}$ are arbitrary constants.
For the relativistic case, the dispersion $\omega_{(\theta) \mathrm{rel}}$ for the localized dilaton becomes massive with a gap $\sqrt{m}$, whereas the dispersion $\omega_{(R) \mathrm{rel}}$ for the localized magnon remains gapless and linear to $|\Vec{k}|$.
For the non-relativistic case, the localized coupled dilaton-magnon mode remains gapless but 
the dispersion relation becomes linear
$\omega\sub{nrel}$ 
from the quadratic (for the undeformed case).

\section{Conclusion and Discussion}
In conclusion, we have considered (quasi-)NG modes 
excited along 
one straight skyrmion line in the relativistic and non-relativistic $\mathbb{C}P^1$ or $O(3)$ sigma models. 
The non-relativistic model describes 
isotropic Heisenberg ferromagnets.
The (quasi-)NG modes in the relativistic model consist of the two translational (Kelvin) modes, 
the localized magnon, and the dilatation mode,  
which are independent of each other and have linear dispersions.
In the non-relativistic model, on the other hand, 
there are the coupled spiral Kelvin wave and localized magnon-dilaton mode with  quadratic dispersions. 
Only when we take into account quasi-NG modes,
the number of gapless modes saturates 
the equality of the Nielsen-Chadha inequality
and satisfies the Watanabe-Brauner's relation,  
in which the commutator between two generators of the internal phase mode and 
the dilatation mode is related to
the topological charge of skyrmions. 
We have also found 
the magnon-dilaton becomes a type-I NG mode 
in the non-relativistic case, 
in the presence of 
the explicit breaking 
terms for the scale symmetry.

Several comments and discussions are addressed here.
The coupled magnon-dilaton found in this paper 
is non-normalizable; 
The effective Lagrangian for that 
is divergent for infinite volume limit $L\to \infty$. 
When there are multiple skyrmion strings, 
one coupled dilaton-magnon is localized on each of them.
While the ``overall" mode, which is a NG mode of 
the global symmetry, 
is non-normalizable,
``relative" modes, which can be regarded as 
locally NG modes for approximate local transformations, 
are normalizable, 
as was shown in Refs.~\cite{Eto:2006db,Eto:2007yv}.

The ${\mathbb C}P^1$ manifold has 
the K\"ahler form $\omega = i du \wedge du^*/(1+|u|^2)^2$
and the topological charge,  
$\pi_2({\mathbb C}P^1) \simeq {\mathbb Z}$,
is the pullback of this form 
into a two-dimensional space perpendicular to 
the skyrmion string. 
Skyrmion strings are admitted in any nonlinear sigma model 
with K\"ahler target spaces $M$ with $\pi_2(M) \neq 0$,
such as the projective space ${\mathbb C}P^N$ and 
the Grassmann sigma model \cite{Eto:2007yv}. 
With a locally defined one-form $\alpha$ satisfying 
$\omega = d\alpha$, the first-order 
time derivative term can be constructed in the Lagrangian, 
and so 
our results can be extended to general K\"ahler manifolds. 

Quantum effects on localized type-II modes 
remain an important problem although they were previously 
studied in a vortex with localized type-II non-Abelian  NG modes 
\cite{Nitta:2013wca}; 
localized type-II NG modes remain gapless, unlike the case of 
relativistic theories in which all NG modes 
in 1+1 dimensions 
are gapped through quantum corrections 
consistent with the Coleman-Mermin-Wagner theorem. 
Quasi-NG modes are in general gapped,  
taking into account quantum corrections even in the bulk 3+1 dimensions, 
because they are not 
associated with an exact symmetry of Lagrangians.
In our case, the magnon-dilaton is a half-genuine NG mode;
therefore, the fate in quantum corrections is a 
non-trivial question.
The analysis of Sec.~\ref{sec:baby} suggests 
that the magnon-dilaton becomes  type-I 
when the dilaton gets a potential term 
from receiving quantum corrections.

In our previous paper \cite{Kobayashi:2014xua}, 
we studied the NG modes of a domain wall 
in the $O(3)$ sigma model with a potential term 
admitting two discrete vacua \cite{Abraham:1992vb}, 
that describes ferromagnets with one easy axis.
A skyrmion studied in this paper 
and a domain wall in the massive $O(3)$ sigma model 
can be related by 
a dimensional reduction \cite{Eto:2006mz},  
as in the case between Yang-Mills instantons and 
BPS magnetic monopoles. 
How type-II NG modes and corresponding 
commutation relations for a skyrmion and a domain wall 
are related to each other 
remains a  problem for future study.  

In $d=3+1$ dimensions, the massive $O(3)$ sigma model admits 
a composite soliton of skyrmion strings ending on a domain wall, 
known as a D-brane soliton 
\cite{Gauntlett:2000de,Isozumi:2004vg}.
D-brane solitons exist also in two-component Bose-Einstein 
condensates \cite{Kasamatsu:2010aq},
for which 
NG modes have been studied in the presence of 
a domain wall 
\cite{Takeuchi:2013mwa,Watanabe:2014zza,Takahashi:2014vua}. 
Investigating NG modes for such a composite soliton will be 
an interesting approach for further study.

\section*{Acknowledgments}
We thank  Rina Takashima and Daisuke Takahashi 
for useful discussions 
and Haruki Watanabe for explaining their previous results 
\cite{Watanabe:2014pea}.
This work is supported in part by Grant-in-Aid for Scientific Research (Grants No. 22740219 (M.K.) and No. 25400268 (M.N.)) and the work of M. N. is also supported in part by the ``Topological Quantum Phenomena" Grant-in-Aid for Scientific Research on Innovative Areas (No. 25103720) from the Ministry of Education, Culture, Sports, Science and Technology (MEXT) of Japan.



\end{document}